\documentclass[dvips,11pt]{article}
\usepackage{amsmath,amssymb,amsthm,graphicx,latexsym}
\newcommand{\doublespacing}{\let\CS=\@currsize\renewcommand{\baselinesstrech}
{2.0}\tiny\CS}

\newcommand{\bd}{\begin{document}}
\newcommand{\ed}{\end{document}}
\newcommand{\bc}{\begin{center}}
\newcommand{\ec}{\end{center}}
\newcommand{\vs}{\vspace}

\begin{document}

\bc {\huge \bf Scattering states of a particle,  } \ec

\vs{.2cm}

\bc {\huge \bf with position-dependent mass, } \ec

\vs{.2cm}

\bc {\huge \bf in a ${\cal{PT}}$ symmetric heterojunction   } \ec

\vs{.5cm}

\bc
{\it \large A. Sinha{\footnote {e-mail : anjana23@rediffmail.com}} \\
Department of Applied Mathematics, \\
Calcutta University, \\
92, A.P.C. Road, Kolkata - 700 009 \\ INDIA } \ec

\vs{1cm}

\begin{abstract}

\noindent The study of a particle with position-dependent
effective mass (pdem), within a double heterojunction is extended
into the complex domain --- when the region within the
heterojunctions is described by a non Hermitian ${\cal{PT}}$
symmetric potential. After obtaining the exact analytical
solutions, the reflection and transmission coefficients are
calculated, and plotted as a function of the energy. It is
observed that at least two of the characteristic features of non
Hermitian ${\cal{PT}}$ symmetric systems --- viz., left / right
asymmetry and anomalous behaviour at spectral singularity, are
preserved even in the presence of pdem. The possibility of charge
conservation is also discussed.

\vs{3cm}


\vs{1cm}


\end{abstract}

\pagebreak

\section{Introduction}

Ever since the pioneering work of Bender et. al. more than a
decade ago, the fact that a class of non-Hermitian Hamiltonians
admit real and discrete spectrum under certain conditions, is well
established \cite{pt-original-1,pt-original-2}. Non-Hermitian
hamiltonians having ${\cal{PT}}$ symmetry ($P \rightarrow $ {\it
parity}, $T \ \rightarrow $ {\it time reversal}) form a special
class in this category, as they admit real and discrete spectrum
for exact ${\cal{PT}}$ symmetry and complex conjugate pairs of
energy when this space-time symmetry is spontaneously broken, the
transition occurring at the so-called exceptional point
\cite{pt-spon1,pt-spon2}. Naturally, numerous attempts have been
made by various scientists to extend the framework of quantum
mechanics into the complex domain \cite{pt-1,pt-2,pt-3}.
Theoretical predictions for ${\cal{PT}}$ symmetric systems exist
in quantum field theory, mathematical, atomic and solid state
physics, classical optics, etc \cite{pt-appl}. A pair of coupled
active LCR circuits --- one with amplification the other with
equivalent attenuation, exhibits ${\cal{PT}}$ symmetry \cite{lcr}.
Of late, experimental confirmation of such non-Hermitian
${\cal{PT}}$ symmetric concepts have been observed in optics, in
${\cal{PT}}$ symmetric crystals with a complex refractive index
distribution $ n(x) = n_0 + n_R (x) + i n_I (x)$, where $n_0$
represents a constant background index, $n_R (x)$ is the real
index profile (even) of the structure, and $n_I (x)$ stands for
the gain or loss component (odd) [10-17]. Unlike ordinary
crystals, complex crystals show unique properties
--- e.g., violation of Fresnel's Law of Bragg scattering, double
refraction, power oscillations, non-reciprocal diffraction,
handedness or left-right asymmetry, anomalous transport,
unidirectional invisibility, etc. In fact, these unique features
together with the occurrence of exceptional points in the discrete
spectrum and spectral singularities in the continuous spectrum,
are characteristic of ${\cal{PT}}$ symmetric non Hermitian
Hamiltonians, unknown to Hermitian ones. A lasing medium embedded
in the spatial region $ |z| < a_0$ where the dielectric constant
satisfies the ${\cal{PT}}$ symmetry condition $ \epsilon
(-\vec{r}) = \epsilon ^* (\vec{r}) $, behaves as a laser
oscillator (LO) for positive  $ {\rm{Im}} \ \epsilon (\vec{r}) $
signifying gain, or as a coherent perfect absorber (CPA) for
negative $ {\rm{Im}} \ \epsilon (\vec{r}) $ representing loss,
provided the dielectric constant is real valued and constant $
\epsilon (\vec{r}) = \epsilon _0 $ (say) outside i.e., for $ |z|
> a_0$. While a LO can emit outgoing coherent waves, a CPA can
fully absorb incoming coherent waves \cite{longhi-pra82}. The
interesting part of non-Hermitian Hamiltonians is that even a
single ${\cal{PT}}$ cell can exhibit unconventional features
\cite{pt-optics2}.

On the other hand the study of quantum mechanical systems with
position dependent effective mass (pdem) has received a boost in
recent times with major developments in nanofabrication techniques
of semiconductor devices [19-26]. The spatial dependence on the
effective mass of the particle arises due to its interaction with
an ensemble of particles within the device, as the particle
propagates from left to right. For example, in
Al$_{x}$Ga$_{1-x}$As, as the mole fraction $x$ varies along the
$z$-axis, so does the effective mass of the charge carrier
(electron or hole). Pdem formalism is extremely important in
describing the electronic and transport properties of quantum
wells and quantum dots, impurities in crystals, He-clusters,
quantum liquids, semiconductor heterostructures, etc. In a recent
work, we obtained the exact analytical scattering solutions of a
particle (electron or hole) in a semiconductor double
heterojunction --- potential well / barrier --- where the
effective mass of the particle varies with position inside the
heterojunctions \cite{epl-anjana}. It was observed that the
spatial dependence on mass within the well / barrier introduces a
nonlinear component in the plane wave solutions of the continuum
states. Additionally, the transmission coefficient increases with
increasing energy, finally approaching unity, whereas the
reflection coefficient follows the reverse trend, going to zero.

This study is presented as a sequel to the work done in ref.
\cite{epl-anjana}, in an attempt to extend the pdem formalism
further into the complex domain. Instances of such attempts are
found in other works as well --- e.g., the problem of relativistic
fermions subject to a ${\cal{PT}}$ symmetric potential in the
presence of position-dependent mass was studied in ref.
\cite{castro}, while exact solutions of Schr\"{o}dinger equation
for ${\cal{PT}}$ / non ${\cal{PT}}$ symmetric and non-Hermitian
Morse and P\"{o}schl-Teller potentials were obtained with pdem by
applying a point canonical transformation method in ref
\cite{schro-pdm}. In the present work, the main stress will be
given on the scattering phenomenon in a ${\cal{PT}}$ symmetric
double heterojunction with pdem --- a special form of
semiconductor device consisting of a thin layer of ${\cal{PT}}$
symmetric material sandwiched between two normal semiconductors,
such that the mass of the charge carrier (electron or hole) varies
with the doping concentration (and hence position) within the
heterojunctions, but is constant outside. In particular, we shall
see what new properties (if any) can be expected from ${\cal{PT}}$
symmetric heterojunctions with pdem, with special emphasis on the
behaviour of the reflection and transmission coefficients \\
~ (i) with respect to the direction of incidence of the particle \\
(ii) at the spectral singularity. \\
Additionally, we shall also derive explicit relations for current
and charge densities, to see whether these are conserved in such a
device. \\
For this purpose, we shall consider a double
heterojunction, with the potential function in the intermediate
region satisfying the ${\cal{PT}}$ symmetry condition $ V(-z) = V
^* (z) $, but assuming a real valued constant outside
\begin{equation}\label{pot-form}
        V = \left\{
    \begin{array}{lcl}
        & & \displaystyle V_R(z) + i V_I (z)
        \  , \ \ \ \ a_1 <  z  < a_2 \\
        & & \displaystyle V_{01} = V_R(a_1) \ \ , \ \ \ - \infty < z < a_1 \\
        & & \displaystyle V_{02} = V_R(a_2) \ \ , \ \ \ \ \ a_2 <  z  < \infty
    \end{array}
        \right.
\end{equation}
where $a_1 , a_2$ represent the heterojunctions. The mass of the
charge carrier is assumed to be of the form
\begin{equation}\label{mass-form}
        m = \left\{
    \begin{array}{lcl}
        & & \displaystyle m(z)
        \qquad \ \ \ \ , \ \ \ \ \ a_1 <  z  < a_2 \\
        & & \displaystyle m_1 = m(a_1)  \ , \ \ \ - \infty < z < a_1 \\
        & & \displaystyle m_2 = m(a_2)  \ , \ \ \ \ \ a_2 < z < \infty
    \end{array}
        \right.
\end{equation}
Thus, the mass $m(z)$ and the real part of the potential function
viz. $V_R (z)$  are considered to be continuous throughout the
semiconductor device. We shall mainly concentrate on obtaining the
exact analytical solutions of the scattering states of a particle
with pdem inside a ${\cal{PT}}$ symmetric double heterojunction,
which is essential to study the nature of the reflection and the
transmission coefficients.

\vs{.2cm}

The article is organized as follows : For the sake of
completeness, the position-dependent-mass Schr\"{o}dinger equation
is introduced in Section 2, and the method of obtaining the
solutions is discussed briefly. To give a better insight into the
physical nature of the problem, we shall study an explicit model
in Section 3, and plot the potential and mass functions as a
function of $z$ in Fig. 1, and the scattering solutions in Fig. 2.
The transmission and reflection coefficients are also calculated,
and their behaviour is discussed with respect to the relative
strengths of the coupling parameters of the potential (both real
and imaginary parts) and the mass functions. Since non-Hermitian
Hamiltonians (with constant mass) are known to show peculiar
behaviour at spectral singularities and also exhibit left-right
asymmetry, the transmission and reflection coefficients are
plotted in Figures 3, 4, 5  and 6, to check whether similar
phenomena are observed in the presence of pdem as well. Section 4
is devoted to the conservation of charge and current densities for
non Hermitian ${\cal{PT}}$ systems with pdem. Finally, Section 5
is kept for Conclusions and Discussions.

\section{Theory}

\noindent We start with the basic assumption that the one
dimensional time independent Schr\"{o}dinger equation associated
with a particle endowed with pdem is the same for Hermitian and
non Hermitian systems, and is given by
\begin{equation}\label{H-em}
\begin{array}{lcl}
    \displaystyle H_{EM} (z) \psi (z)  & \equiv & \left[ T_{EM} (z) +
    V(z) \right]
    \psi (z)  =  E \psi (z) \\ \\
    \qquad V(z) & = & \displaystyle V_R (z) + i V_I (z) \\
\end{array}
\end{equation}
in the intermediate region within the heterojunctions. $T_{EM}$ is
the kinetic energy term given by \cite{pdm4,harrison}
\begin{equation}\label{T-em}
    \begin{array}{lcl}
    T_{EM} &=& \displaystyle \frac{1}{4} \left( m^{\alpha} p
    m^{\beta} p m ^{\gamma} + m^{\gamma} p
    m^{\beta} p m ^{\alpha} \right) \\ \\
    &=& \displaystyle \frac{1}{2} p \left( \frac{1}{m}
    \right) p
    \end{array}
\end{equation}
where $ p = \displaystyle - i \hbar \frac{d}{dz} $ is the momentum
operator. It is to be noted here that the kinetic energy term is
considered to be Hermitian. The non Hermiticity is introduced
through the potential term $V(z)$, with an even real part $V_R(z)$
and an odd imaginary part $V_I(z)$. The ambiguity parameters
$\alpha \ , \ \beta \ , \ \gamma $ obey the von Roos constraint
\cite{pdm4}
    \begin{equation}\label{abg}
        \alpha + \beta + \gamma = -1
    \end{equation}
In the absence of a unique or universal choice for the ambiguity
parameters, several suggestions exist in literature
\cite{benDaniel-Duke,z-k,m-m,l-k}, etc. However, for continuity
conditions at the abrupt interfaces, and well behaved ground state
energy \cite{marrow, thomsen}, we shall restrict ourselves to the
BenDaniel-Duke choice, viz., $ \alpha = \gamma = 0 \ , \ \beta =
-1 $. Incidentally, this particular choice consistently produces
the best fit to experimental results \cite{proceed}. Furthermore,
we shall work in units $\hbar = c = 1$, and use prime to denote
differentiation w.r.t. $z$. Thus, inside the potential well $a_1 <
z < a_2$, the Hamiltonian for the particle with pdem reduces to
\cite{plastino}
\begin{equation}\label{h-in}
        H = \displaystyle - \frac{1}{2m(z)} \frac{d^2}{dz^2} - \left(
        \frac{1}{2m(z)} \right) ^{\prime} \frac{d}{dz} + V_R(z) +
        i V_I (z)
\end{equation}
whereas, outside the well, $ z < a_1 $ and $ z > a_2 $, the
particle obeys the conventional Schr\"{o}dinger equation :
\begin{equation}\label{sch-out}
    \displaystyle \left\{ - \frac{1}{2m_{1,2}} \frac{d^2}{dz^2} +
    V_{01,02}
    \right\} \psi (z)  = E \psi (z)
\end{equation}
having plane wave solutions. In case we consider a wave incident
from left, the solutions in the two regions are
\begin{equation}\label{psi-out}
\begin{array}{lcl}
    \psi _L (z)  &=& \displaystyle e^{i k_1 z} + R e^{-ik_1 z} \ , \ - \infty <
    z < a_1 \\ \\
    \psi _R (z)  &=& T e^{i k_2 z} \ , \ \qquad \qquad a_2 <
    z < \infty \\
\end{array}
\end{equation}
where $R$ and $T$ denote the reflection and transmission
amplitudes, and
\begin{equation}\label{k}
    k_{1,2} = \displaystyle  \sqrt{ 2 m_{1,2} \left( E - V_{01,02} \right) }
\end{equation}

\vs{.1cm}

\noindent To find the solution in the region $a_1 < z < a_2$, we
make use of the following transformations \cite{br-pr}
    \begin{equation}\label{psi-z}
         \psi _{in} (z) = \displaystyle \left\{ 2 m(z) \right\} ^{1/4} \phi
         (\rho) \qquad , \qquad \rho = \displaystyle \int \sqrt{2 m(z)} dz
    \end{equation}
which reduce the Schr\"{o}dinger equation for pdem, to one for
constant mass, viz.,
\begin{equation}\label{schro-const-m}
    \displaystyle -  \frac{d^2 \phi }{d \rho ^2} + \left\{
    \widetilde{V} (\rho) - E \right\} \phi  = 0
\end{equation}
with
\begin{equation}\label{v-tilde}
    \widetilde{V} (\rho) = \displaystyle V(z) + \frac{7}{32} \frac{m^{\prime
    \ 2}}{m^3} - \frac{m^{\prime \prime}}{8 m^2}
\end{equation}
Some definite practical forms of $V(z)$ and $m(z)$ give exact
analytical solutions of (\ref{schro-const-m}). An explicit example
in the next section illustrates our purpose.

\section{Explicit model : ${\cal{PT}}$ symmetric potential well with position dependent mass}

\noindent We consider the following ansatz for the potential,
whose real part describes a diffused quantum well
\begin{equation}\label{pot-1}
        V(z) = \left\{
    \begin{array}{lcl}
        & & \displaystyle - \ \frac{\mu _1 }{1 + z^2} + i
        \frac{\mu _2 z}{1+z^2}
        \ \  , \ | z | < a_0   \\ \\
        & & \displaystyle - \frac{\mu _1 }{1 + a_0^2} \ = \
        V_0 \qquad \ , \ | z | > a_0  \\
    \end{array}
    \right.
\end{equation}
Let the mass of the particle be
\begin{equation}\label{mass-1}
        m(z) = \left\{
    \begin{array}{lcl}
        & & \displaystyle \frac{\beta ^2}{2 \left(1 + z^2 \right)}
        \ \qquad \  , \ | z | < a_0 \\ \\
        & & \displaystyle \frac{\beta ^2}{2 \left(1 + a_0^2 \right)} = m_0 \
        , \ | z | > a_0 \\
    \end{array}
        \right.
\end{equation}
where $\mu _1 , \ \mu _2 , \ \beta$ are some constant parameters.

\noindent For a better understanding of the mass dependence and
the potential in the semiconductor device, we plot $m(z)$ and the
real and imaginary parts of $V(z) $ as a function of $z$ in Fig.
1, for a suitable set of parameter values, viz., $\beta = 4 , \
\mu = 3, \ a_0 = 4 $.

{\begin{figure}[hp]
\begin{center}
\scalebox{0.5}{\includegraphics{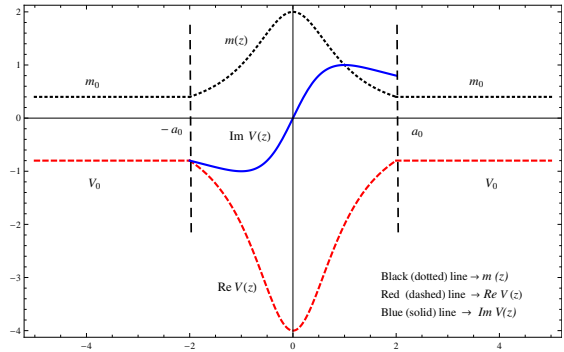}}
\label*{}\caption{\small {Colour online : Plot showing $m(z)$ and
$V(z)$ w.r.t. $z$ }}
\end{center}
\end{figure}}

\noindent For the spatial mass dependence given by eq.
(\ref{mass-1}), eq. (\ref{psi-z}) transforms the coordinate $z$ to
\begin{equation}\label{rho}
    \rho = \beta \sinh ^{-1} z
\end{equation}
so that after some straightforward algebra $\widetilde{V} (\rho)$
in eq. (\ref{v-tilde}) reduces to
\begin{equation}\label{v-sech}
    \widetilde{V} (\rho) = \displaystyle \frac{1}{4 \beta ^2} -
    \frac{V_1}{\beta ^2} {\rm{sech}} ^2  \frac{\rho}{\beta} + i
    \frac{V_2}{\beta ^2} {\rm{sech}} \  \frac{\rho}{\beta} \ \tanh
    \frac{\rho}{\beta}
\end{equation}
Thus equation (\ref{schro-const-m}) may be written as
\begin{equation}\label{schro-rho}
    \displaystyle \frac{d^2 \phi}{d \bar{\rho} ^2} + \left \{ \kappa ^2 +
    V_1 {\rm{sech}} ^2 \bar{\rho} - i V_2 {\rm{sech}} \ \bar{\rho} \
    \tanh \bar{\rho} \right\} \phi = 0
\end{equation}
\begin{equation}\label{k}
    {\rm{where}} \qquad \displaystyle \kappa ^2 =  E \beta ^2 - \frac{1}{4 }
    \ \ , \ \ \displaystyle \bar{\rho} = \frac{\rho}{\beta}
\end{equation}
and the parameters $V_1 ,  V_2 $ depend on the constants $\mu _1 ,
\mu _2 $ and $\beta$, through the equations
\begin{equation}\label{lambda}
    V_1 = \displaystyle  \mu _1 \beta ^2 + \frac{1}{4} \qquad ,
    \qquad V_2 = \mu _2 \beta ^2
\end{equation}

\noindent Let us introduce a new variable
\begin{equation}\label{y}
    y = \displaystyle  \frac{1+i \sinh  \bar{\rho}}{2}
\end{equation}
and write the solutions of (\ref{schro-rho}) as
\begin{equation}\label{phi-u}
    \phi = \displaystyle y^p (1-y)^q \ u(y)
\end{equation}
In terms of the new variable $y$, equation (\ref{schro-rho})
reduces to the hypergeometric equation
\begin{equation}\label{y-u}
    \begin{array}{lll}
    \displaystyle y(1-y) \frac{d^2 u}{dy^2} &+& \left\{ \left(
    2p  + \displaystyle \frac{1}{2} \right)
    -  \left( 2p +2q + 1 \right) y \right\} \displaystyle \frac{du}{dy}
    \\ \\
    &-& \displaystyle \left\{
    \kappa ^2 - (p+q) ^2 \right\} u = 0
    \end{array}
\end{equation}
where
\begin{equation}\label{p-q}
    \begin{array}{lcl}
    p &=& \displaystyle \frac{1}{2} \pm \frac{1}{2}
    \sqrt{\frac{1}{4} + V_1 - V_2} \\ \\
    q &=& \displaystyle \frac{1}{2} \pm \frac{1}{2}
    \sqrt{\frac{1}{4} + V_1 + V_2}
    \end{array}
\end{equation}
Now, (\ref{y-u}) has complete solution \cite{flugge}
\begin{equation}\label{u-hypergeometric}
    u = \displaystyle P \ _2F_1 \left( a,b,c; y
    \right) + \displaystyle Q y^{1-c} \ _2F_1 \left( 1+a-c,1+b-c,2-c; y
    \right)
\end{equation}
where $P$ and $Q$ are constants, and the parameters $a$ and $b$
are as defined below :
\begin{equation}\label{ab}
    a = \displaystyle p + q - i \kappa
    \ \ , \ \ b = \displaystyle p + q + i \kappa
\end{equation}

\noindent After some straightforward algebra, the final solution
to the pdem Schr\"{o}dinger equation (\ref{H-em}), within the
potential well $ | z | < a_0$, is obtained as
\begin{equation}\label{psi-in}
    \begin{array}{lll}
    \psi _{in} (z) &=& \displaystyle \frac{\beta ^{1/2}}{2 ^{p+q}}
    \ \left( 1+ i z \right) ^{p - 1/4} \left( 1- i z \right) ^{q - 1/4}
    \ \left\{ P \ _2F_1 \left( a,b,c; y \right)  \right. \\ \\
    & & \displaystyle + \ Q \left( \frac{1-iz}{2} \right) ^{1-c}
    \ _2F_1 \left( 1+a-c, 1+b-c ,2-c; y \left. \right) \right\}
    \end{array}
\end{equation}
where $ y = (1+iz)/2$. Outside the well ($ | z | >  a_0 $), the
solutions are given by eq (\ref{psi-out}), with $ k_1 = k_2$.

\noindent The solution in the entire region is plotted in Fig. 2
for the same set of parameter values as in Fig. 1, viz., $\beta =
4 , \ \mu = 3, \ a_0 = 4 $. With the help of Mathematica, the
constants $P, Q$ and the reflection and transmission amplitudes
$|R|$ and $|T|$ respectively, are determined using modified
boundary conditions for pdem systems
\cite{benDaniel-Duke,boundary} --- \\
the functions $\displaystyle \psi (z) $ and $\displaystyle
\frac{1}{m(z)} \frac{d \psi (z)}{dz} $ should be continuous at
each heterojunction $\pm a_0$. That the effect of pdem is to
introduce a non-linear component in the solution is obvious from
the figure. This finding is analogous to the Hermitian model with
pdem \cite{epl-anjana}.

{\begin{figure}[hp]
\begin{center}
\scalebox{0.4}{\includegraphics{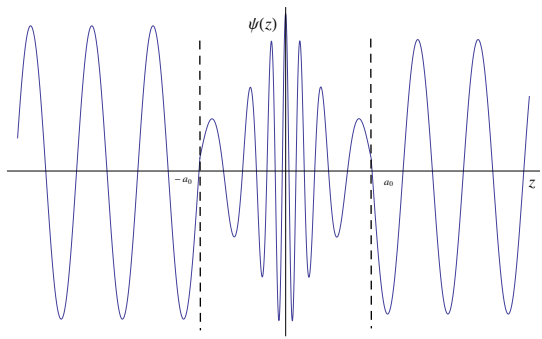}}
\label*{}\caption{\small {Colour online : A plot of Re $ \psi (z)$
vs $z$; Dashed (black) lines show the abrupt heterojunctions }}
\end{center}
\end{figure}}

\noindent For real and discrete spectrum, $ \displaystyle | V_2| <
V_1 + 1/4 $, implying
\begin{equation}\label{mu1-mu2-ep}
    \displaystyle | \mu _2 | \ < \ \mu _1 + \frac{1}{2 \beta ^2}
\end{equation}
However, in this work we are interested in the scattering states
only, i.e., positive $ \kappa ^2 $ . Hence
    $$ E > \displaystyle \frac{1}{4 \beta ^2} $$

{\begin{figure}[hp]
\begin{center}
\scalebox{0.8}{\includegraphics{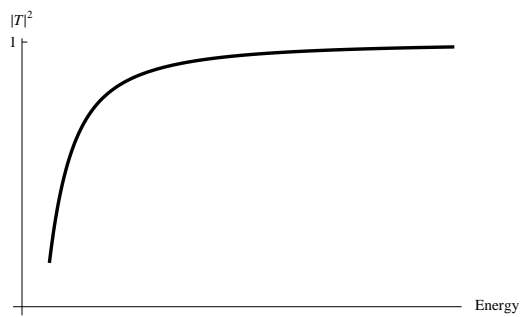}}
\label*{}\caption{\small {Plot of $|T|^2$ vs $E$ --- same for left
and right incidence }}
\end{center}
\end{figure}}

\noindent The transmission and reflection amplitudes $T$ and $R$
respectively, are plotted in Figures 3, 4 and 5, for both left and
right incidence, with the help of Mathematica. While $|T|$ comes
out to be the same for either case, the reflection coefficient
depends on whether the particle enters from left or right. $|R|$
is normal ($|R| < 1$) when the particle enters from left --- the
absorptive side ($ {\rm {Im}} \ V(z) < 0$), and anomalous ($|R| >
1$) when the particle enters from right
--- the emissive side ($ {\rm {Im}} \ V(z) > 0$). Thus, this phenomenon of
left-right asymmetry, characteristic of non hermitian ${\cal{PT}}$
symmetric potentials with constant mass particles
\cite{zafar1,zafar2, cannata-annals}, remains unaltered even when
the particle mass has a spatial dependence.

{\begin{figure}[hp]
\begin{center}
\scalebox{0.8}{\includegraphics{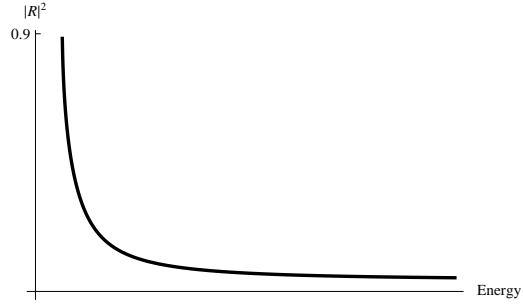}}
\label*{}\caption{\small {Plot of $|R|^2$ vs $E$ for left
incidence  }}
\end{center}
\end{figure}}

{\begin{figure}[hp]
\begin{center}
\scalebox{0.8}{\includegraphics{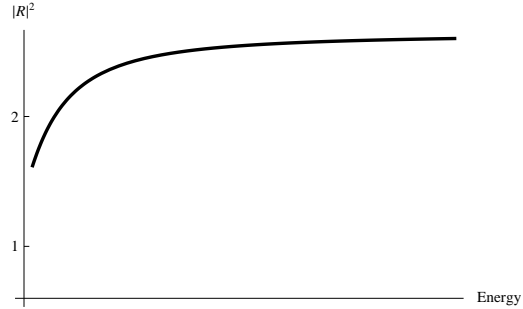}}
\label*{}\caption{\small {Plot of $|R|^2$ vs $E$ for right
incidence }}
\end{center}
\end{figure}}

\noindent Another interesting feature worth discussing here is
when the well is replaced by a barrier --- i.e., $ \displaystyle
\mu _1 < 0 $, so that
\begin{equation}\label{v-rho-ss}
    \displaystyle \tilde{V}( \bar{\rho}) = \displaystyle  \frac{1}{4} +
    \left( \mu _1 \beta ^2 - \frac{1}{4} \right) {\rm{sech}}^2 \ \bar{\rho}
    \displaystyle + \ i \mu _2 \beta ^2 {\rm{sech}} \ \bar{\rho} \
    \tanh \bar{\rho}
\end{equation}

\noindent For particles with constant mass, the potential in
(\ref{v-rho-ss}) is known to admit a spectral singularity, with
the reflection and transmission coefficients blowing up at the
positive energy \cite{mostafazadeh-ss, zafar-ss}
\begin{equation}\label{ss}
    E_s = \displaystyle \frac{1}{4} \left[ \mid V_2 \mid - \left(
    \frac{1}{4} + V_1 \right) \right]
\end{equation}
\begin{equation}\label{condition-ss}
    {\rm{when}} \qquad    \displaystyle \mid V_2  \mid \ > \
    \mid V_1 \mid \ + \ \frac{{\rm{sign \ of
    \ }} V_1}{4}
\end{equation}
For the pdem non Hermitian heterojunction considered here, the
condition (\ref{condition-ss}) comes out to be
\begin{equation}\label{condition-new}
    \displaystyle \mu _2 \ > \ \mu _1 + \frac{1}{2 \beta ^2}
\end{equation}
so that spectral singularity occurs at
\begin{equation}\label{E-ss}
    \displaystyle E_s = \displaystyle \frac{1}{4} \left( \mu _2 -
    \mu _1 \right) \beta ^2 - \frac{1}{8}
\end{equation}

\noindent Interestingly, even in a ${\cal{PT}}$ symmetric double
heterojunction, with a spatially varying mass, both $|R|^2$ and
$|T|^2$ blow up at the value of $E_s$ given in eq. (\ref{E-ss}),
as observed in Fig. 6.

\pagebreak

{\begin{figure}[hp]
\begin{center}
\scalebox{0.6}{\includegraphics{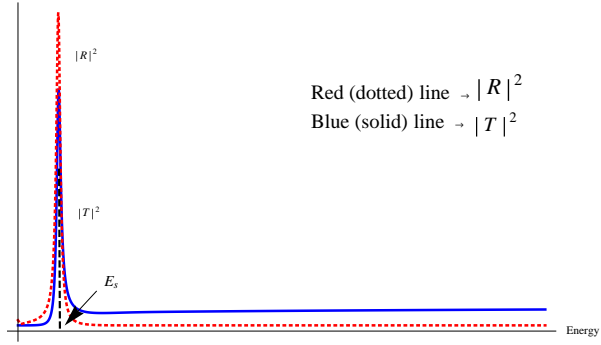}}
\label*{}\caption{\small {Colour online : Plot of $|R|^2$ and
$|T|^2$ vs $Energy$ for ${\cal{PT}}$ symmetric barrier : $E_s$
shows the spectral singularity }}
\end{center}
\end{figure}}

\section{Current and Charge Conservation}

Since we are dealing with non Hermitian ${\cal{PT}}$ symmetric
systems, it would be interesting to check whether the current and
charge densities are conserved in this case. The boundary
conditions used in this work are
\begin{equation}\label{boundary}
    \displaystyle \psi (z) | _L = \psi (z) | _R \qquad , \qquad
    \displaystyle \frac{1}{m(z)} \frac{d \psi }{dz} | _L =
    \displaystyle \frac{1}{m(z)} \frac{d \psi }{dz} | _R
\end{equation}
where $L$ and $R$ stand for left and right side of the
heterojunctions. This choice of the boundary conditions ensures
that current and charge densities are conserved in the Hermitian
system \cite{benDaniel-Duke,boundary}. In this section we explore
the possibility of this choice for the non Hermitian case
discussed here. It is known from earlier works that charge density
(say $\omega$) and current density (say $\mathbf{j}$) for non
Hermitian quantum systems with constant mass, obey the equation of
continuity (for exact or unbroken ${\cal{PT}}$ symmetry)
\cite{sinha-roy-FTC}
\begin{equation}\label{pt-cont}
    \displaystyle \frac{\partial \omega}{\partial t} +
    \mathbf{\nabla} _{\rho} \cdot \mathbf{j} = 0
\end{equation}
only if the current and charge densities are redefined as
\begin{equation}\label{current-charge}
    \omega = \displaystyle \phi ^* \eta \phi \qquad , \qquad
    \mathbf{j} = \displaystyle i \left( \frac{d \phi ^* }{d \rho} \eta
    \phi - \phi ^* \eta \frac{d \phi}{d \rho} \right)
\end{equation}
where $\eta$ is a linear, invertible, Hermitian operator, with
respect to which the non Hermitian Hamiltonian $H$ is pseudo
Hermitian  :
\begin{equation}\label{eta-pseudo}
    H^{\dagger} = \eta ^{-1} H \eta
\end{equation}
The interesting point to note here is that $\eta$ does not have a
unique representation \cite{mostafazadeh-JMP}. For ${\cal{PT}}$
symmetric potentials consisting of an even real part $V_R$, and an
odd imaginary part $V_I$, viz.,
\begin{equation}\label{v-pt}
    V (z) = V_R (z) + i V_I (z) \qquad {\rm{or}} \qquad
    \widetilde{V}(\rho) = V_R (\rho) + i V_I (\rho)
\end{equation}
$\eta $ may be represented by the parity operator ${\cal{P}}$
\cite{mostafazadeh-JMP,ahmed-PLA}. For the non Hermitian
${\cal{PT}}$ Scarf II potential, viz., $V(\rho) = \displaystyle -
V_R \ {\rm{sech}}^2 \rho - i V_I \ {\rm{sech}} \ \rho \ \tanh \rho
$ some definite forms of $\eta$ are given in ref.
\cite{bb-rr-PLA}. With $\eta = {\cal{P}}$, and the forms given in
ref. \cite{bb-rr-PLA}, it is easy to check that for exact or
unbroken ${\cal{PT}}$ symmetry, the equation of continuity is
obeyed in the transformed coordinate system $\rho$, where the
problem reduces to one with constant mass, viz. eq.
(\ref{pt-cont}). Now, if one applies the inverse transformations
of those in eq. (\ref{psi-z}), one can verify by straightforward
algebra that eq. (\ref{pt-cont}) in the transformed coordinate
system $\rho$ (with constant mass) can be mapped to a similar
equation of continuity in the original $z$-coordinate system
(where the mass of the particle is dependent on its position) :
\begin{equation}\label{cont-pdm}
    \displaystyle \frac{\partial \bar{\omega}}{\partial t} +
    \mathbf{\nabla} _{z} \cdot \mathbf{\bar{j}} = 0
\end{equation}
provided the charge density ($\bar{\omega}$) and current density
($\mathbf{\bar{j}}$) in the original space are mapped to those
($\omega$ and $\mathbf{j}$ respectively) in the transformed space
by
\begin{equation}\label{w-j}
\begin{array}{lll}
    \bar{\omega} &=& \displaystyle \frac{1}{\sqrt{2m(z)}} \ \omega
    = \displaystyle \psi ^* \eta \psi \\
    \mathbf{\bar{j}} &=& \displaystyle \frac{1}{\sqrt{2m(z)}} \
    \mathbf{j} = \displaystyle \frac{i}{\sqrt{2m(z)}}
    \left( \frac{d \psi ^*}{dz} \eta \psi - \psi ^* \eta
    \frac{d \psi}{dz} \right) \\
\end{array}
\end{equation}
Thus, if the charge and current densities ($ \bar{\omega}$ and
$\mathbf{\bar{j}}$ respectively) for non Hermitian ${\cal{PT}}$
symmetric quantum systems with position dependent effective mass
are given a modified definition in accordance with eq.
(\ref{w-j}), then the boundary conditions in eq. (\ref{boundary})
ensure conservation of current so long as ${\cal{PT}}$ symmetry is
unbroken or exact. However, with the spontaneous breakdown of this
space-time  symmetry, the current is no longer conserved.

\section{Conclusions and Discussions}

To conclude, in this work we studied a special form of
semiconductor device consisting of a thin layer of ${\cal{PT}}$
symmetric material sandwiched between two normal semiconductors,
such that the mass of the charge carrier (electron or hole) varies
with position within the heterojunctions, but is constant outside.
The mass $m(z)$ and the real part of the potential $V_R(z)$ are
taken to be continuous throughout the device. We obtained the
exact analytical solutions for the scattering states of a particle
inside such a semiconductor device and also the reflection and
transmission amplitudes, $R$ and $T$ respectively. Additionally,
we also obtained explicit relations for current and charge
densities for such a ${\cal{PT}}$ symmetric double heterojunction.

The primary aim of this work was to extend the pdem formalism into
the complex domain, to see if the spatial dependence of mass
introduces any new feature in case of non Hermitian ${\cal{PT}}$
symmetric double heterojunctions. It is observed that at least two
of the general features of ${\cal{PT}}$ symmetric potentials ---
viz., left-right asymmetry and blowing up of reflection and
transmission coefficients at a spectral singularity --- are
preserved even for particles with pdem. The effect of the pdem is
simply to introduce a non-linear component in the otherwise plane
wave solution, within the heterojunctions. Another interesting
feature we discussed here is the equation of continuity. For non
Hermitian ${\cal{PT}}$ quantum systems with pdem, a modified
definition of charge and current densities as given in eq.
(\ref{w-j}) renders the conservation of current for exact
${\cal{PT}}$ symmetry. To the best of our knowledge, this is a new
observation.

Treating particles with position dependent effective mass because
of varying doping concentration in semiconductor devices of
extremely small dimensions, is found to give better explanation to
experimentally observed phenomena. On the other hand ${\cal{PT}}$
symmetric waveguides fabricated from iron-doped LiNbO$_3$ are also
a reality \cite{pt-optics4,nature-kottos}. The time may not be far
when a new generation of sophisticated, integrated devices are
constructed based on the simple model discussed in this work. So
our next attempt would be to see if the results of this study are
typical of this particular example, or model-independent.

\section{Acknowledgement}

Financial assistance for the work was provided for by the
Department of Science and Technology, Govt. of India, through its
grant SR/WOS-A/PS-06/2008. The author thanks Prof. R. Roychoudhury
for a careful reading of the manuscript. Thanks are also due to
the unknown referees for some interesting suggestions.

\end{document}